\documentclass[aps,preprint]{revtex4}%
\usepackage{amsfonts}
\usepackage{amsmath}
\usepackage{amssymb}
\usepackage{graphicx}%
\begin{document}
\title{{\LARGE SIMPLE DERIVATION OF SCHWARZSCHILD, LENSE-THIRRING,
\ REISSNER-NORDSTR\"{O}M, KERR AND KERR-NEWMAN METRICS}}
\author{Marcelo Samuel Berman$^{1}$ }
\affiliation{$^{1}$Instituto Albert Einstein \ - Av. Candido de Abreu, 469 - \ \# 1503 \ -
Ed. Sobral Pinto, Centro C\'{\i}vico. 80730-440 - Curitiba - PR - Brazil}
\begin{abstract}
An effort has been made, in order to derive without "hard" mathematics, the
forms of \ SCHWARZSCHILD, LENSE-THIRRING, REISSNER-NORDSTR\"{O}M, KERR and
KERR-NEWMAN metrics.

\end{abstract}
\maketitle

\begin{center}
{\LARGE SIMPLE DERIVATION OF SCHWARZSCHILD, LENSE-THIRRING,
REISSNER-NORDSTR\"{O}M, KERR AND KERR-NEWMAN METRICS}

{\LARGE \bigskip}

Marcelo Samuel Berman

\end{center}

\bigskip

\bigskip{\Large I. Introduction}

\bigskip Shortly after the appearance of Einstein's General Relativistic field
equations, the first static and spherically symmetric solution became
available: it was Schwarzschild's metric (Schwarzschild, 1916). It described
the gravitational field around a point like mass \ \ $M$\ \ . Afterwards, the
first rotational metric was developed: Lense-Thirring solution (Thirring and
Lense, 1918). It described the field around a rotating sphere at the origin.
Nevertheless, it was only an approximate solution, that only represented a
weak field, in the slow rotation case. Reissner and Nordstr\"{o}m's metric
(Reissner, 1916), generalized Schwarzschild's by charging the mass source in
the origin. It was only in the sixties of last century, that a rigorous
solution representing the rotation of a central mass, was discovered by Roy
Kerr, in what is now called Kerr's metric (Kerr, 1963). Immediately
afterwards, the generalization to a charged rotating central mass was
supplied, which is now called Kerr-Newman's metric (Newman et al. 1965).

\bigskip

The literature on black holes is, by now, very extensive. In recent times,
some elementary books have appeared, which intend to teach beginners without
using thoroughly, its natural tool, i.e., tensor calculus. (For instance:
Taylor and Wheeler, 2000; or, Raine and Thomas, 2005). Nevertheless, \ it has
been lacking a simple derivation of any of those metrics, without the use of
sophisticated mathematics. Taylor and Wheeler (2000) refer to an article about
the impossibility of a simple derivation of the Schwarzschild's metric (Gruber
et al., 1988). While preparing a new textbook (Berman, to appear), I have made
some elementary derivations, now being presented here.

\bigskip{\Large II. Simple Derivation of Schwarzschild's metric}

\bigskip From Special Relativity, we know that in the absence of gravitation,
the metric is given by:

\bigskip$ds^{2}=c^{2}dt^{2}-d\sigma^{2}=c^{2}dt^{2}-dx^{2}-dy^{2}-dz^{2}%
=c^{2}dt^{2}-dr^{2}-r^{2}d\Omega$ \ \ \ \ \ \ \ \ \ \ \ \ ,\ \ \ \ \ \ \ \ \ (II.1)

\bigskip

where,

\bigskip

$d\Omega=d\theta^{2}+\sin^{2}\theta$ $\ d\phi^{2}$ \ \ \ \ \ \ \ \ .\ \ \ \ \ \ \ \ \ \ \ \ \ \ \ \ \ \ \ \ \ \ \ \ \ \ \ \ \ \ \ \ \ \ \ \ \ \ \ \ \ \ \ \ \ \ \ \ \ \ \ \ \ \ \ \ \ \ \ \ \ \ \ \ \ \ \ \ \ \ \ \ \ \ \ \ \ \ \ \ \ \ (II.2)

\bigskip

In the above, ($x,y,z$) and ($r,\theta,\phi$) are Cartesian and spherical
coordinates, respectively.

\bigskip

When a gravitational field appears, the metric is "curved", but in the static,
spherically symmetric case, we can write generally that:

\bigskip

$ds^{2}=g_{00}c^{2}dt^{2}-g_{rr}dr^{2}-r^{2}d\Omega$ \ \ \ \ \ \ \ \ \ \ \ \ \ \ ,\ \ \ \ \ \ \ \ \ \ \ \ \ \ \ \ \ \ \ \ \ \ \ \ \ \ \ \ \ \ \ \ \ \ \ \ \ \ \ \ \ \ \ \ \ \ \ \ \ \ \ \ \ \ \ \ \ \ \ \ \ \ \ (II.3)

\bigskip

where, \ $g_{00}$\ \ and \ \ $g_{rr}$\ \ should be functions depending on the
mass \ $M$\ \ and the radial coordinate \ $r$\ \ .

\bigskip

In order to estimate the first function, we consider a fixed event
($dr=d\theta=d\phi=0$). In this case, the proper time is given by
\ \ $ds^{2}\equiv d\tau^{2}$\ :

\bigskip

$d\tau^{2}=g_{00}c^{2}dt^{2}$\ \ \ \ \ \ \ \ \ . \ \ \ \ \ \ \ \ \ \ \ \ \ \ \ \ \ \ \ \ \ \ \ \ \ \ \ \ \ \ \ \ \ \ \ \ \ \ \ \ \ \ \ \ \ \ \ \ \ \ \ \ \ \ \ \ \ \ \ \ \ \ \ \ \ \ \ \ \ \ \ \ \ \ \ \ \ \ \ \ \ \ \ \ \ \ \ \ \ \ \ \ \ \ (II.4)\ 

\bigskip

In the above, \ \ $dt$\ \ \ represents coordinate \ time, i.e., the time far
from the central mass distribution (call it, the time for a clock at
infinity). From the Principle of Equivalence, we know that the relation
between coordinate time, and the proper time, as measured close to a mass
distribution, is given by (Sexl and Sexl, 1979)\ :

\bigskip

$d\tau\cong\frac{dt}{\left[  1+\frac{GM}{c^{2}R}\right]  }=\frac{dt}{\left[
1-\frac{\Delta U}{c^{2}}\right]  }$ \ \ \ \ \ \ \ \ \ \ \ \ \ \ \ \ \ \ \ \ . \ \ \ \ \ \ \ \ \ \ \ \ \ \ \ \ \ \ \ \ \ \ \ \ \ \ \ \ \ \ \ \ \ \ \ \ \ \ \ \ \ \ \ \ \ \ \ \ \ \ \ \ \ \ \ \ \ \ \ \ \ \ \ \ \ \ (II.5)

\bigskip

On squaring the last expression, we find:

\bigskip

$d\tau^{2}\cong\left[  1-\frac{2GM}{c^{2}R}\right]  c^{2}dt^{2}$%
\ \ \ \ \ \ \ \ . \ \ \ \ \ \ \ \ \ \ \ \ \ \ \ \ \ \ \ \ \ \ \ \ \ \ \ \ \ \ \ \ \ \ \ \ \ \ \ \ \ \ \ \ \ \ \ \ \ \ \ \ \ \ \ \ \ \ \ \ \ \ \ \ \ \ \ \ \ \ \ \ \ \ \ \ \ \ \ \ \ (II.6)

\bigskip

\bigskip Consider a free-fall experiment. A particle falls, and has velocities
\ $v_{A}$\ \ \ and \ $v_{B}$\ \ , while passing through points \ $A$\ \ and
\ $B$\ \ . The free-fall happens due to a gravitational field; the Newtonian
conservation law can be applied, yielding:

\bigskip

$\frac{E}{M}=\frac{1}{2}v_{A}^{2}+U_{A}=\frac{1}{2}v_{B}^{2}+U_{B}%
$\ \ \ \ \ \ \ \ \ \ \ \ , \ \ \ \ \ \ \ \ \ \ \ \ \ \ \ \ \ \ \ \ \ \ \ \ \ \ \ \ \ \ \ \ \ \ \ \ \ \ \ \ \ \ \ \ \ \ \ \ \ \ \ \ \ \ \ \ \ \ \ \ \ \ \ \ \ \ \ (II.6a)

\bigskip

which can be expressed as:

\bigskip

$\frac{1}{2}\left(  \frac{v_{A}^{2}}{c^{2}}-\frac{v_{B}^{2}}{c^{2}}\right)
=-\frac{\Delta U}{c^{2}}$ \ \ \ \ \ \ \ \ \ \ \ \ \ \ \ \ \ \ \ . \ \ \ \ \ \ \ \ \ \ \ \ \ \ \ \ \ \ \ \ \ \ \ \ \ \ \ \ \ \ \ \ \ \ \ \ \ \ \ \ \ \ \ \ \ \ \ \ \ \ \ \ \ \ \ \ \ \ \ \ \ \ \ \ \ \ \ \ \ \ \ (II.6b)

\bigskip

From Special Relativity, if \ $\tau$\ \ represents proper time, we can write
for points \ $A$\ \ and \ \ $B$\ \ when the particle passes in times
\ \ $t_{A}$\ \ and \ $t_{B}$\ :

\bigskip

$\frac{t_{B}}{t_{A}}=\frac{\tau\text{ \ }\sqrt{1-\frac{v_{B}^{2}}{c^{2}}}%
}{\tau\text{ \ }\sqrt{1-\frac{v_{A}^{2}}{c^{2}}}}\cong1+\frac{1}{2}\left(
\frac{v_{A}^{2}}{c^{2}}-\frac{v_{B}^{2}}{c^{2}}\right)  =-\frac{\Delta
U}{c^{2}}$ \ \ \ \ \ \ \ \ \ \ \ . \ \ \ \ \ \ \ \ \ \ \ \ \ \ \ \ \ \ \ \ \ \ \ \ \ \ \ \ \ \ \ \ \ \ \ \ \ \ \ \ \ \ \ \ \ \ \ (II.6c)\ \ 

\bigskip

It is to be noticed, that the gravitational field being weak, we used the approximation:

\bigskip

$\left(  1+\Delta\right)  ^{n}\cong1+n\Delta$ \ \ \ \ \ \ \ \ \ \ \ \ \ \ . \ \ \ \ \ \ \ \ \ \ \ \ \ \ \ \ \ \ \ \ \ \ \ \ \ \ \ \ \ \ \ \ \ \ \ \ \ \ \ \ \ \ \ \ \ \ \ \ \ \ \ \ \ \ \ \ \ \ \ \ \ \ \ \ \ \ \ \ \ \ \ \ \ \ \ \ \ \ \ \ \ \ (II.7)

\bigskip

From (II.6c), above, the derivation of (II.5) becomes evident.

\bigskip

The reader can check, that we now have obtained \ $g_{00}$\ :

\bigskip

$g_{00}\cong\left[  1-\frac{2GM}{c^{2}R}\right]  $\ \ \ \ \ \ \ \ . \ \ \ \ \ \ \ \ \ \ \ \ \ \ \ \ \ \ \ \ \ \ \ \ \ \ \ \ \ \ \ \ \ \ \ \ \ \ \ \ \ \ \ \ \ \ \ \ \ \ \ \ \ \ \ \ \ \ \ \ \ \ \ \ \ \ \ \ \ \ \ \ \ \ \ \ \ \ \ \ \ \ \ \ \ \ \ \ \ \ (II.8)\ 

\bigskip

\bigskip In Sexl and Sexl (1979), however, the reader will find that it is not
possible to derive the corresponding change in coordinate length, from proper
length, without delving into elaborated mathematics \ (Taylor and Wheeler,
2000; Gruber et all, 1988).

\bigskip

On remembering that a weak gravitational field does not differ much from its
Minkowski metric counterpart, as shown in (II.1), we make the hypothesis that
the determinant of the metric tensor, which in this case is diagonal, is
approximately Minkowskian, and, thus, we shall solve the problem, because we have:

\bigskip

$g=-1,$ \ \ \ \ \ \ \ \ \ \ \ \ \ \ \ \ \ in Cartesian coordinates, or,

\bigskip

$g=-R^{4}\sin^{2}\theta$ \ \ , \ \ \ \ \ in spherical coordinates.

\bigskip

In our case, this implies in that:

\bigskip

$g=g_{00}g_{RR}g_{\theta\theta}g_{\phi\phi}\cong-R^{4}\sin^{2}\theta$ \ \ ,
\ \ \ \ \ \ \ \ in spherical coordinates.

\bigskip

We justify, because the sought metric must be Lorentzian at infinity, and this
entails relation,

\bigskip

\bigskip$g_{00}g_{RR}=1$ \ \ \ \ \ \ \ .

As we have already found \ $g_{00}$\ \ , we now have at our disposal the result:

\bigskip

$g_{RR}=\left(  g_{00}\right)  ^{-1}\cong\left[  1-\frac{2GM}{c^{2}R}\right]
^{-1}$\ \ \ \ \ \ \ \ \ \ \ \ \ \ \ \ \ \ ,\ \ \ \ \ \ \ \ \ \ \ \ \ \ \ \ \ \ \ \ \ \ \ \ \ \ \ \ \ \ \ \ \ \ \ \ \ \ \ \ \ \ \ \ \ \ \ \ \ \ \ \ \ \ \ \ \ \ \ \ \ \ \ \ \ (II.9)\ 

\bigskip

and , unaltered, \ 

$\bigskip$

$d\Omega=d\theta^{2}+\sin^{2}\theta$ $\ d\phi^{2}$ \ \ \ \ \ \ \ \ .\ \ \ \ \ \ \ \ \ \ \ \ \ \ \ \ \ \ \ \ \ \ \ \ \ \ \ \ \ \ \ \ \ \ \ \ \ \ \ \ \ \ \ \ \ \ \ \ \ \ \ \ \ \ \ \ \ \ \ \ \ \ \ \ \ \ \ \ \ \ \ \ \ \ \ \ \ \ \ \ \ \ \ \ \ \ \ (II.10)

\bigskip

The last one, implies that ,

\bigskip

$g_{\theta\theta}=R^{2}$\ \ \ \ \ \ \ \ \ ,

\bigskip

and, \ \ \ \ 

$\bigskip$

$g_{\phi\phi}=R^{2}\sin^{2}\theta$\ \ .\ \ 

\bigskip

Though we derived an approximation, our result is, in fact, exact; the
Schwarzschild's metric is given, then, by:

\bigskip

$ds^{2}=\left[  1-\frac{2GM}{c^{2}R}\right]  c^{2}dt^{2}-\left[  1-\frac
{2GM}{c^{2}R}\right]  ^{-1}dr^{2}-r^{2}d\Omega$
\ \ \ \ \ \ \ \ \ \ \ \ \ \ ,\ \ \ \ \ \ \ \ \ \ \ \ \ \ \ \ \ \ \ \ \ \ \ \ \ \ \ \ \ \ \ \ \ \ \ \ (II.11)\bigskip

\bigskip

When the field is indeed weak, we write the above in the following form:

\bigskip

\bigskip$ds^{2}\cong\left[  1-\frac{2GM}{c^{2}R}\right]  c^{2}dt^{2}-\left[
1+\frac{2GM}{c^{2}R}\right]  dr^{2}-r^{2}d\Omega$
\ \ \ \ \ \ \ \ \ \ \ \ \ \ \ ,\ \ \ \ \ \ \ \ \ \ \ \ \ \ \ \ \ \ \ \ \ \ \ \ \ \ \ \ \ \ \ \ \ \ \ \ \ (II.12)\bigskip

\bigskip It is a pity that only about a century after its first derivation, we
could find a solution for it without tensor calculus.

\bigskip

[ It came to my attention a different "simplified derivation" of
Schwarzschild's metric, published by Rabinowitz (2006), based on a paper by
Winterberg (2002). Their trick, was taught by me since 1989, when I began to
lecture General Relativity; nevertheless, when I tried to publish in a
Journal, it was rejected because, according to the referee, there was a
mixture between radial and transverse concepts, what was "wrong". The American
Journal of Physics, received an appeal by me, saying that in equation (II.6a),
what matters is the magnitude of the velocity, not the direction; this was in
the year 2001, but the Journal\ stood by the referee's report, and my paper
was rejected. The derivation in the present communication, is different from
the other one].

\bigskip

{\Large III. Isotropic form of Schwarzschild's metric}

\bigskip

It is desirable that Schwarzschild's metric be cast in the isotropic form,
which is meant by:

\bigskip

$ds^{2}=g_{00}c^{2}dt^{2}-g_{\sigma\sigma}$ $d\sigma^{2}$
\ \ \ \ \ \ \ \ \ \ \ \ \ \ . \ \ \ \ \ \ \ \ \ \ \ \ \ \ \ \ \ \ \ \ \ \ \ \ \ \ \ \ \ \ \ \ \ \ \ \ \ \ \ \ \ \ \ \ \ \ \ \ \ \ \ \ \ \ \ \ \ \ \ \ \ \ \ \ \ \ \ \ \ \ \ \ \ \ (III.1)

\bigskip

In order to find the correct isotropic form, we imagine that we make a change
in coordinates, from \ $R$\ \ to $\rho$\ , and that we wish to find the
relation between both, so that, when we begin with the standard
Schwarzschild's metric \ (II.11), we find the isotropic metric:

\bigskip

$ds^{2}\cong\left[  1-\frac{2GM}{c^{2}\rho}\right]  c^{2}dt^{2}-\left[
1+\frac{2GM}{c^{2}\rho}\right]  $ $d\sigma^{2}$ \ \ \ \ \ \ \ \ \ \ \ \ \ \ ,\ \ \ \ \ \ \ \ \ \ \ \ \ \ \ \ \ \ \ \ \ \ \ \ \ \ \ \ \ \ \ \ \ \ \ \ \ \ \ \ \ \ \ \ \ \ \ \ \ \ \ \ (III.2)\ \ 

\bigskip

with,

\bigskip

$d\sigma^{2}=d\rho^{2}+\rho^{2}d\Omega$
\ \ \ \ \ \ \ \ \ \ \ \ \ \ \ \ \ \ \ . \ \ \ \ \ \ \ \ \ \ \ \ \ \ \ \ \ \ \ \ \ \ \ \ \ \ \ \ \ \ \ \ \ \ \ \ \ \ \ \ \ \ \ \ \ \ \ \ \ \ \ \ \ \ \ \ \ \ \ \ \ \ \ \ \ \ \ \ \ \ \ \ \ \ \ \ \ \ \ \ (III.3)

\bigskip

We took the \ $g_{00}=$\ $g_{00}(\rho)$\ \ \ to be the same function as
\ $g_{00}(R)$\ \ ; it could work or not. In fact, it works.

\bigskip

We go right to the solution of the problem:

\bigskip

$R\cong\left[  1+\frac{2GM}{c^{2}\rho}\right]  ^{\frac{1}{2}}\rho\cong\left[
1+\frac{GM}{c^{2}\rho}\right]  \rho\cong\rho+\frac{GM}{c^{2}}$
\ \ \ \ \ \ \ \ \ \ \ \ \ \ \ . \ \ \ \ \ \ \ \ \ \ \ \ \ \ \ \ \ \ \ \ \ \ \ \ \ \ \ \ \ \ \ \ \ \ \ \ \ \ \ \ \ \ \ \ (III.4)

\bigskip

With the above substitution, in the metric (III.2), we obtain,

\bigskip

$ds^{2}=\left[  1-\frac{2GM}{c^{2}\rho}\right]  c^{2}dt^{2}-\left[
1+\frac{2GM}{c^{2}\rho}\right]  \left[  d\rho^{2}+\rho^{2}d\Omega\right]  $
\ \ \ \ \ \ \ \ \ \ \ \ \ \ .\ \ \ \ \ \ \ \ \ \ \ \ \ \ \ \ \ \ \ \ \ \ \ \ \ \ \ \ \ \ \ \ \ \ \ \ (III.5)\bigskip

\bigskip In the same level of approximation, the last form of the metric, is
indistinguishable from the following one, which is the exact isotropic form of
Schwarzschild's metric:

\bigskip

$ds^{2}=\frac{\left[  1-\frac{GM}{2c^{2}\rho}\right]  ^{2}}{\left[
1+\frac{GM}{2c^{2}\rho}\right]  ^{2}}c^{2}dt^{2}-\left[  1+\frac{GM}%
{2c^{2}\rho}\right]  ^{4}\left[  d\rho^{2}+\rho^{2}d\Omega\right]  $
\ \ \ \ \ \ \ \ \ \ \ \ \ \ ,\ \ \ \ \ \ \ \ \ \ \ \ \ \ \ \ \ \ \ \ \ \ \ \ \ \ \ \ \ \ \ \ \ \ \ \ (III.6)\bigskip

\bigskip

{\Large IV. Simple derivation of Lense-Thirring metric}

\bigskip

\bigskip For a rotating central mass, we start first with the approximate
isotropic metric of last Section (relation III.5):

\bigskip$ds^{2}=\left[  1-\frac{2GM}{c^{2}\rho}\right]  c^{2}dt^{2}-\left[
1+\frac{2GM}{c^{2}\rho}\right]  \left[  d\rho^{2}+\rho^{2}d\Omega\right]  $
\ \ \ \ \ \ \ \ \ \ \ \ \ \ .\ \ \ \ \ \ \ \ \ \ \ \ \ \ \ \ \ \ \ \ \ \ \ \ \ \ \ \ \ \ \ \ \ \ \ \ (IV.1)\bigskip

Consider now a transformation from the above spherical coordinates, \ \ $\rho$
, $\theta$ , $\phi$\ , \ to a rotating frame, defined by the new coordinates
\ $R$\ \ , \ $\theta$\ \ , \ $\tilde{\phi}$\ , whereby:

\bigskip

$R=\rho$ \ \ \ \ \ , \ \ \ \ \ \ \ \ \ \ \ \ \ \ \ \ \ \ \ \ \ \ \ \ \ \ \ \ \ \ \ \ \ \ \ \ \ \ \ \ \ \ \ \ \ \ \ \ \ \ \ \ \ \ \ \ \ \ \ \ \ \ \ \ \ \ \ \ \ \ \ \ \ \ \ \ \ \ \ \ \ \ \ \ \ \ \ \ \ \ \ \ \ \ \ \ \ \ \ \ \ \ \ \ \ \ \ \ \ (IV.2a)

\bigskip

$\tilde{\phi}=\phi-\omega$ $t$\ \ \ \ \ \ \ , \ \ \ \ \ \ \ \ \ \ \ \ \ \ \ \ \ \ \ \ \ \ \ \ \ \ \ \ \ \ \ \ \ \ \ \ \ \ \ \ \ \ \ \ \ \ \ \ \ \ \ \ \ \ \ \ \ \ \ \ \ \ \ \ \ \ \ \ \ \ \ \ \ \ \ \ \ \ \ \ \ \ \ \ \ \ \ \ \ \ \ \ \ \ \ \ \ \ \ \ (IV.2b)

\bigskip

$d\tilde{\phi}=d\phi-\omega$ $dt$\ \ \ \ . \ \ \ \ \ \ \ \ \ \ \ \ \ \ \ \ \ \ \ \ \ \ \ \ \ \ \ \ \ \ \ \ \ \ \ \ \ \ \ \ \ \ \ \ \ \ \ \ \ \ \ \ \ \ \ \ \ \ \ \ \ \ \ \ \ \ \ \ \ \ \ \ \ \ \ \ \ \ \ \ \ \ \ \ \ \ \ \ \ \ \ \ \ \ \ \ \ (IV.2c)

\bigskip

\bigskip

The new expression for the metric, will be:

\bigskip

$ds^{2}=\left[  1-2U-\left(  1+2U\right)  \omega^{2}R^{2}\sin^{2}%
\theta\right]  c^{2}dt^{2}-\left(  1+2U\right)  d\sigma^{2}+2\left(
1+2U\right)  \omega R^{2}\sin^{2}\theta$ $d\phi$ $dt$ \ \ \ , (IV.3)

\bigskip

where,

\bigskip

$U=\frac{GM}{c^{2}R}$ \ \ \ \ \ \ \ \ \ \ . \ \ \ \ \ \ \ \ \ \ \ \ \ \ \ \ \ \ \ \ \ \ \ \ \ \ \ \ \ \ \ \ \ \ \ \ \ \ \ \ \ \ \ \ \ \ \ \ \ \ \ \ \ \ \ \ \ \ \ \ \ \ \ \ \ \ \ \ \ \ \ \ \ \ \ \ \ \ \ \ \ \ \ \ \ \ \ \ \ \ \ \ \ \ \ \ \ \ \ \ \ \ (IV.4)

\bigskip

Note that we have dropped the \ tilde \ from \ \ $\phi$\ \ .

\bigskip

\bigskip

Consider now the greatest difference between the last metric and the
non-rotating one, i.e., the existence of a non-diagonal metric element,

\bigskip

$2\left(  1+2U\right)  \omega R^{2}\sin^{2}\theta$ $d\phi$ $dt$ \ \ \ \ \ \ . \ \ \ \ \ \ \ \ \ \ \ \ \ \ \ \ \ \ \ \ \ \ \ \ \ \ \ \ \ \ \ \ \ \ \ \ \ \ \ \ \ \ \ \ \ \ \ \ \ \ \ \ \ \ \ \ \ \ \ \ \ \ \ \ \ \ \ \ \ \ \ \ \ \ \ \ \ \ (IV.5)

\bigskip

\bigskip

We can define a Newtonian angular momentum \ $J$\ \ , so that:

\bigskip

$2\left(  1+2U\right)  \omega R^{2}\sin^{2}\theta$ $d\phi$ $dt=2\left(
1+2U\right)  \frac{J}{M}$ $d\phi$ $dt$ \ \ \ \ \ \ \ . \ \ \ \ \ \ \ \ \ \ \ \ \ \ \ \ \ \ \ \ \ \ \ \ \ \ \ \ \ \ \ \ \ \ \ \ \ \ \ \ \ \ \ \ (IV.6)\ 

\bigskip

\bigskip

It is easy to check that we have employed a natural definition for \ $J$\ \ ,
in the above equation. As \ $U$\ \ and \ $J$ \ , are small, so that the
rotating metric is very approximately similar to the non-rotating one, we may
also write:

\bigskip

\ $g_{\phi t}$ $d\phi$ $dt=2\left(  1+2U\right)  \omega R^{2}\sin^{2}\theta$
$d\phi$ $dt=\left(  1+2U\right)  \frac{J}{M}$ $d\phi$ $dt\cong\left[
\frac{2J}{M}+\frac{4GJ}{R}\right]  $ $d\phi$ $dt$ \ \ \ \ . \ \ \ \ (IV.7)\ 

\bigskip

\bigskip

By the same token, the extra term in \ $g_{00}$\ , is given by the product\ of
\ $\omega$\ \ with the non-diagonal metric coefficient \ \ $g_{\phi t}$ \ \ , i.e.,

\bigskip

$\omega\left[  \frac{J}{M}+\frac{2GJ}{R}\right]  $ \ \ \ \ \ \ \ , \ \ \ \ \ \ \ \ \ \ \ \ \ \ \ \ \ \ \ \ \ \ \ \ \ \ \ \ \ \ \ \ \ \ \ \ \ \ \ \ \ \ \ \ \ \ \ \ \ \ \ \ \ \ \ \ \ \ \ \ \ \ \ \ \ \ \ \ \ \ \ \ \ \ \ \ \ \ \ \ \ \ \ \ \ \ \ \ \ \ \ \ \ \ \ \ \ (IV.8)

\bigskip

which can be neglected, for being a second order infinitesimal.

\bigskip

\bigskip

The above results constitute the Lense-Thirring metric, which we now have
shown to be derived with simple mathematics.

\bigskip

\bigskip

\bigskip

{\Large V. Simple derivation of Reissner-Nordstr\"{o}ms metric}

\bigskip

\bigskip Consider now a statical spherically symmetric metric, representing a
charged mass \ $M$\ \ . We keep the same requirement adopted in order to
obtain Schwarzschild's metric, because it is still the same argument:

\bigskip

$g=g_{00}g_{RR}g_{\theta\theta}g_{\phi\phi}=-R^{4}\sin^{2}\theta$
\ \ \ \ \ \ \ \ (as in Minkowski's metric) \ \ . \ \ \ \ \ \ \ \ \ \ \ \ \ \ \ \ \ \ \ \ \ \ \ (V.1)

\bigskip

\bigskip As the part in \ $d\Omega$\ \ is to be kept intact, we are going to
write down a standard metric form, with \ $g_{00}=\left(  g_{RR}\right)
^{-1}$\ ; \ because the given metric has to reduce to Schwarzschild's one, in
case \ $Q=0$\ \ , where \ \ $Q$ \ is the electric charge, we may write:

\bigskip

$g_{00}\cong\left[  1-\frac{2GM}{c^{2}R}+kQ^{n}R^{m}c^{s}G^{p}\right]
$\ \ \ \ \ \ \ \ \ \ \ \ \ \ . \ \ \ \ \ \ \ \ \ \ \ \ \ \ \ \ \ \ \ \ \ \ \ \ \ \ \ \ \ \ \ \ \ \ \ \ \ \ \ \ \ \ \ \ \ \ \ \ \ \ \ \ \ \ \ \ \ \ (V.2)\ 

\bigskip

The third term in the r.h.s. above, was written with the understanding that
only charge, radial distance, and the constants $c$\ \ \ \ and \ \ $G$\ \ ,
can have any influence in the sought metric. There is a point that makes the
electric case different from the gravitational part: the result should not
change when a positive charge is substituted by a negative one. We impose
then, that \ $n$\ \ is even; it should also be positive, so that increasing
the charge, will increase the change in the metric , when compared with the
Schwarzschild's one, for \ $k>0$\ .\ From simplicity arguments, we would like
to choose the smallest positive and even number: \ $n=2$\ .\ \ One more thing:
the relative dependence of the constants, in the \ $Q$\ \ term, must be
similar to the ones in the \ $M$\ \ term: this makes us impose that:

\bigskip

$m=-n$ \ \ \ \ \ \ \ \ \ \ \ \ \ , \ \ \ \ \ \ \ \ \ \ \ \ \ \ \ \ \ \ \ \ \ \ \ \ \ \ \ \ \ \ \ \ \ \ \ \ \ \ \ \ \ \ \ \ \ \ \ \ \ \ \ \ \ \ \ \ \ \ \ \ \ \ \ \ \ \ \ \ \ \ \ \ \ \ \ \ \ \ \ \ \ \ \ \ \ \ \ \ \ \ \ \ \ \ \ \ \ (V.3)

\bigskip

$s=-2n$\ \ \ \ \ \ \ \ \ \ \ \ \ . \ \ \ \ \ \ \ \ \ \ \ \ \ \ \ \ \ \ \ \ \ \ \ \ \ \ \ \ \ \ \ \ \ \ \ \ \ \ \ \ \ \ \ \ \ \ \ \ \ \ \ \ \ \ \ \ \ \ \ \ \ \ \ \ \ \ \ \ \ \ \ \ \ \ \ \ \ \ \ \ \ \ \ \ \ \ \ \ \ \ \ \ \ \ \ \ \ (V.4)

\bigskip

If \ $k$\ \ is a pure number, and because the whole term is also,
dimensionally speaking, another pure number, \ we must impose altogether,

\bigskip

$p=1$\ \ \ \ \ \ \ \ \ \ \ \ \ \ \ \ . \ \ \ \ \ \ \ \ \ \ \ \ \ \ \ \ \ \ \ \ \ \ \ \ \ \ \ \ \ \ \ \ \ \ \ \ \ \ \ \ \ \ \ \ \ \ \ \ \ \ \ \ \ \ \ \ \ \ \ \ \ \ \ \ \ \ \ \ \ \ \ \ \ \ \ \ \ \ \ \ \ \ \ \ \ \ \ \ \ \ \ \ \ \ \ \ \ (V.5)\ 

\bigskip

The Reissner-Nordstr\"{o}m temporal metric coefficient, is now in the form:

\bigskip

$g_{00}\cong\left[  1-\frac{2GM}{c^{2}R}+kQ^{2}R^{-2}c^{-4}G\right]
$\ \ \ \ \ \ \ \ \ \ \ \ \ \ . \ \ \ \ \ \ \ \ \ \ \ \ \ \ \ \ \ \ \ \ \ \ \ \ \ \ \ \ \ \ \ \ \ \ \ \ \ \ \ \ \ \ \ \ \ \ \ \ \ \ \ \ \ \ \ \ \ \ (V.6)\ 

\bigskip

We choose now \ \ $k=1$\ \ , because if we would reverse the calculation, and
obtain Einstein's tensor \ \ $G_{\mu\nu}$\ \ from the given metric, we would
find that it would be equal to \ $\kappa T_{\mu\nu}$\ \ , only if the energy
momentum tensor would be constituted by the electric field component, and this
would imply \ $k=1$\ \ .

\bigskip

We now write our final result, which, in fact, is not only approximate, but
indeed exact:

\bigskip

$ds^{2}=\left[  1-\frac{2GM}{c^{2}R}+\frac{GQ^{2}}{c^{4}R^{2}}\right]
c^{2}dt^{2}-\left[  1-\frac{2GM}{c^{2}R}+\frac{GQ^{2}}{c^{4}R^{2}}\right]
^{-1}dR^{2}-R^{2}d\Omega$\ \ \ \ \ \ . \ \ \ \ \ \ \ \ \ \ \ \ \ \ (V.7)\ \ \ \ \ \ \ \ \ \ \ 

\bigskip

{\Large VI. Simple "derivation" of Kerr's metric}

\bigskip

We "derive" here, from Lense-Thirring approximate metric, (referring to a
rotating black hole and in the slow rotating case), the general case of a
rotating mass metric. The derivation goes from the approximate case, towards
the correct generalization; the ultimate recognition, that our derivation is
correct, lies in intricate mathematical calculations, which we will not
present here; we direct the reader to Adler et al. (1975), for the exact derivation.

\bigskip

From Section IV, we may write \ a rotating metric (Lense-Thirring), finding:

\bigskip

$ds^{2}=\left(  1-\frac{2m}{\tilde{\rho}}\right)  dt^{2}-\left(  1+\frac
{2m}{\tilde{\rho}}\right)  d\sigma^{2}-\frac{4ma}{\tilde{\rho}}\sin^{2}\theta$
$d\phi$ $dt$ \ \ \ \ \ \ \ \ , \ \ \ \ \ \ \ \ \ \ \ \ \ \ \ \ \ \ \ \ \ \ \ \ \ \ \ \ \ \ \ \ \ \ \ (VI.1)

\bigskip

where \ \ $m=\frac{GM}{c^{2}}$ \ \ . Notice that in some places, we make
\ $c=1$\ \ .

\bigskip

The reader can check, that, in the approximation \ $\frac{m}{\tilde{\rho}}%
<<1$\ \ , which characterizes L.T. metric, the above expression is equivalent
also to:

\bigskip

$ds^{2}=\frac{\left(  1-\frac{m}{2\tilde{\rho}}\right)  ^{2}}{\left(
1+\frac{m}{2\tilde{\rho}}\right)  ^{2}}dt^{2}-\left(  1+\frac{m}{2\tilde{\rho
}}\right)  ^{4}d\sigma^{2}-\frac{4ma}{\tilde{\rho}\left(  1+\frac{m}%
{2\tilde{\rho}}\right)  }\sin^{2}\theta$ $d\phi$ $dt$ \ \ \ \ \ \ \ \ . \ \ \ \ \ \ \ \ \ \ \ \ \ \ \ \ \ \ \ \ \ \ \ \ \ \ \ \ \ \ \ (VI.2)

\bigskip

This is essentially what we are looking for, in the isotropic form. We now go
to standard form. The reader can check that the desired form is:

\bigskip

$ds^{2}=\left(  1-\frac{2m}{\rho}\right)  dt^{2}-\left(  1-\frac{2m}{\rho
}\right)  ^{-1}d\rho^{2}-\rho^{2}(d\theta^{2}+\sin^{2}\theta$ $d\phi
^{2})-\frac{4ma}{\rho}$ $\sin^{2}\theta$ $d\phi$ $dt$ \ \ . \ \ \ \ \ \ \ (VI.3)

\bigskip

\bigskip The above relation being valid for \ \ $\rho^{2}>>a^{2}$\ \ , it
could be derived, by imposing this approximation, from the exact relation:

\bigskip

$ds^{2}=\frac{\Delta}{\rho^{2}}\left(  dt-a\sin^{2}\theta d\phi\right)
^{2}-\frac{\sin^{2}\theta}{\rho^{2}}\left[  \left(  r^{2}+a^{2}\right)
d\phi-adt\right]  ^{2}-\rho^{2}\left[  \frac{dr^{2}}{\Delta}+d\theta
^{2}\right]  $ \ \ \ \ , \ \ \ \ \ \ \ \ \ \ \ \ (VI.4)

\bigskip

where \ $\Delta$\ , \ $\rho$\ , and \ $a$\ \ \ are defined by:

\bigskip

$\Delta\equiv r^{2}-2mr+a^{2}$ \ \ \ \ \ \ \ , \ \ \ \ \ \ \ \ \ \ \ \ \ \ \ \ \ \ \ \ \ \ \ \ \ \ \ \ \ \ \ \ \ \ \ \ \ \ \ \ \ \ \ \ \ \ \ \ \ \ \ \ \ \ \ \ \ \ \ \ \ \ \ \ \ \ \ \ \ \ \ \ \ \ \ \ \ \ \ \ \ \ \ \ \ \ \ \ \ \ (VI.5)

\bigskip

$\rho^{2}\equiv r^{2}+a^{2}\cos^{2}\theta$ \ \ \ \ \ \ \ \ \ \ \ . \ \ \ \ \ \ \ \ \ \ \ \ \ \ \ \ \ \ \ \ \ \ \ \ \ \ \ \ \ \ \ \ \ \ \ \ \ \ \ \ \ \ \ \ \ \ \ \ \ \ \ \ \ \ \ \ \ \ \ \ \ \ \ \ \ \ \ \ \ \ \ \ \ \ \ \ \ \ \ \ \ \ \ \ \ \ (VI.6)

\bigskip

$a^{2}\equiv\frac{J^{2}}{M^{2}}$ \ \ \ \ \ \ \ \ \ \ \ \ \ \ \ \ \ \ \ \ \ \ . \ \ \ \ \ \ \ \ \ \ \ \ \ \ \ \ \ \ \ \ \ \ \ \ \ \ \ \ \ \ \ \ \ \ \ \ \ \ \ \ \ \ \ \ \ \ \ \ \ \ \ \ \ \ \ \ \ \ \ \ \ \ \ \ \ \ \ \ \ \ \ \ \ \ \ \ \ \ \ \ \ \ \ \ \ \ \ \ \ \ (VI.7)

\bigskip

\bigskip The Kerr metric above is given in Boyer-Lindquist form.

\bigskip

We note again that we have induced and not derived the correct generalization
of L.T. metric into (VI.4), which is valid for any value of the rotation parameter.

\bigskip

{\Large VII. Simple "derivation" of Kerr-Newman metric}

\bigskip

\bigskip

We recall the derivation of Lense-Thirring metric as above: the most general
black hole is characterized by the "exact" rotating metric with mass $M$,
electric charge $Q$ and rotational parameter "$a$"\ and is given by
Kerr-Newman's metric, where in quasi-Cartesian \ form, is given by (Newman et
al., 1965):

\bigskip

$ds^{2}=dt^{2}-dx^{2}-dy^{2}-dz^{2}-\frac{2\left[  M-\frac{Q^{2}}{2r_{0}%
}\right]  r_{0}^{3}}{r_{0}^{4}+a^{2}z^{2}}\cdot F^{2}$ \ \ \ \ \ \ \ \ \ \ \ \ \ \ \ \ \ \ \ \ \ \ \ \ \ \ \ \ \ \ \ \ \ \ \ \ \ \ \ \ \ \ \ \ \ \ \ \ \ \ \ \ \ \ (VII.1)

$F=dt+\frac{Z}{r_{0}}dz+\frac{r_{0}}{\left(  r_{0}^{2}+a^{2}\right)  }\left(
xdx+ydy\right)  +\frac{a\left(  xdy-ydx\right)  }{a^{2}+r_{0}^{2}}$\ \ \ \ \ \ \ \ \ \ \ \ \ \ \ \ \ \ \ \ \ \ \ \ \ \ \ \ \ \ \ \ \ \ \ \ \ \ \ \ \ \ \ \ \ \ \ \ \ \ \ \ \ (VII.2)

$r_{0}^{4}-\left(  r^{2}-a^{2}\right)  r_{0}^{2}-a^{2}z^{2}=0$\ \ \ \ \ \ \ \ \ \ \ \ \ \ \ \ \ \ \ \ \ \ \ \ \ \ \ \ \ \ \ \ \ \ \ \ \ \ \ \ \ \ \ \ \ \ \ \ \ \ \ \ \ \ \ \ \ \ \ \ \ \ \ \ \ \ \ \ \ \ \ \ \ \ \ \ \ \ \ \ \ \ \ \ \ (VII.3)

and

$r^{2}\equiv x^{2}+y^{2}+z^{2}$ \ \ \ \ \ \ \ \ \ \ \ \ \ \ \ \ \ \ \ \ \ \ \ \ \ \ \ \ \ \ \ \ \ \ \ \ \ \ \ \ \ \ \ \ \ \ \ \ \ \ \ \ \ \ \ \ \ \ \ \ \ \ \ \ \ \ \ \ \ \ \ \ \ \ \ \ \ \ \ \ \ \ \ \ \ \ \ \ \ \ \ \ \ \ \ \ \ \ \ (VII.4)

\bigskip

We derive the above result, by writing the Kerr metric in Boyer-Lindquist coordinates,

\bigskip

$ds^{2}=\frac{\Delta}{\rho^{2}}\left(  dt-a\sin^{2}\theta d\phi\right)
^{2}-\frac{\sin^{2}\theta}{\rho^{2}}\left[  \left(  r^{2}+a^{2}\right)
d\phi-adt\right]  ^{2}-\rho^{2}\left[  \frac{dr^{2}}{\Delta}+d\theta
^{2}\right]  $ \ \ \ \ , \ \ \ \ \ \ \ \ (VII.5)

\bigskip

where,

\bigskip

$\Delta\equiv r^{2}-2mr+a^{2}$ \ \ \ \ \ \ \ , \ \ \ \ \ \ \ \ \ \ \ \ \ \ \ \ \ \ \ \ \ \ \ \ \ \ \ \ \ \ \ \ \ \ \ \ \ \ \ \ \ \ \ \ \ \ \ \ \ \ \ \ \ \ \ \ \ \ \ \ \ \ \ \ \ \ \ \ \ \ \ \ \ \ \ \ \ \ \ \ \ \ \ \ \ \ \ \ (VII.6)

\bigskip

$\rho^{2}\equiv r^{2}+a^{2}\cos^{2}\theta$ \ \ \ \ \ \ \ \ \ \ \ . \ \ \ \ \ \ \ \ \ \ \ \ \ \ \ \ \ \ \ \ \ \ \ \ \ \ \ \ \ \ \ \ \ \ \ \ \ \ \ \ \ \ \ \ \ \ \ \ \ \ \ \ \ \ \ \ \ \ \ \ \ \ \ \ \ \ \ \ \ \ \ \ \ \ \ \ \ \ \ \ \ \ \ (VII.7)

\bigskip

The limiting cases of Kerr metric are:

\bigskip

\bigskip

\textbf{A)} Schwarzschild's metric: \ \ \ \ \ \ \ \ \ \ \ \ \ \ \ \ \ we
recover this metric in the limit \ \ $a\rightarrow0$\ \ .

\bigskip

\textbf{B) }Minkowski's metric: \ \ \ \ \ \ \ \ \ \ \ \ \ \ \ \ \ \ \ \ \ \ we
recover when \ \ $m\rightarrow0$\ \ \ and \ \ \ \ $a\rightarrow0$\ \ .

\bigskip

\textbf{C)} Minkowski's rotating Universe: \ \ \ \ \ \ when \ \ $m\rightarrow
0$\ \ \ but \ \ \ $a\neq0$ \ \ .

\bigskip

\textbf{D)} Lense-Thirring metric:
\ \ \ \ \ \ \ \ \ \ \ \ \ \ \ \ \ \ \ \ when \ $a^{2}<<1$\ \ .\ 

\bigskip

\bigskip

In order to check the limiting case of a Minkowski's rotating metric, we may
proceed afresh like it follows. We write:

\bigskip

$ds^{2}=dt^{2}-dx^{2}-dy^{2}-dz^{2}$ \ \ \ \ \ \ \ \ \ \ \ \ \ .

\bigskip

In cylindrical coordinates, the above metric would become:

\bigskip

$d\tilde{s}^{2}=dt^{2}-\left[  d\tilde{r}^{2}+\tilde{r}^{2}d\tilde{\phi}%
^{2}+d\tilde{z}^{2}\right]  $ \ \ \ \ \ \ \ \ \ \ \ \ \ \ .

\bigskip

We revert to a rotating metric by means of the transformation of coordinates below:

\bigskip

$r=\tilde{r}$ \ \ \ \ \ \ \ \ \ \ ,

$z=\tilde{z}$\ \ \ \ \ \ \ \ \ \ \ ,

$t=\tilde{t}$\ \ \ \ \ \ \ \ \ \ \ ,

\bigskip

and,

$\bigskip$

$d\phi=d$ $\tilde{\phi}+\omega$ $.d$ $\tilde{t}$ \ \ \ \ \ \ \ \ \ \ \ \ .

\bigskip

We find now:

\bigskip

$ds^{2}=-\left[  dr^{2}+r^{2}d\phi^{2}+dz^{2}\right]  +2r^{2}\omega$ $d\phi$
$dt+\left[  c^{2}-\omega^{2}r^{2}\right]  $ $dt^{2}$ \ \ \ \ \ \ \ \ \ \ .

\bigskip

The existence of a crossed term \ $g_{\phi t}\neq0$\ , is characteristic
feature of this model, likewise the Kerr metric (see Lense-Thirring metric in
Section IV).\ \ \ It is usual to demand that rotating metrics must have
non-diagonal metric tensor components, an example being given by our calculations.

\bigskip

\bigskip

\textbf{We now extend Kerr metric into K.N. metric by the addition of a term
such that:}

\bigskip

\bigskip

\textbf{a)} when \ $Q=0$ \ ,
\ \ \ \ \ \ \ \ \ \ \ \ \ \ \ \ \ \ \ \ \ \ \ \ \ \ we obtain \ \ Kerr metric;

\textbf{b)} if \ \ $Q\neq0$\ \ \ but \ \ $a=0$\ \ , \ \ \ \ \ \ \ \ \ \ \ we
obtain Reissner-Nordstr\"{o}m's metric\ .

\bigskip

By the same token under which we showed how to obtain
\ Reissner-Nordstr\"{o}m's metric, from Schwarzschild's, we redefine here the
\ $\Delta$ \ , writing:

\bigskip

$\Delta\equiv r^{2}-2mr+a^{2}+\frac{GQ^{2}}{c^{4}}$ \ \ \ \ \ \ \ . \ \ \ \ \ \ \ \ \ \ \ \ \ \ \ \ \ \ \ \ \ \ \ \ \ \ \ \ \ \ \ \ \ \ \ \ \ \ \ \ \ \ \ \ \ \ \ \ \ \ \ \ \ \ \ \ \ \ \ \ \ \ \ \ \ \ \ \ \ \ \ \ \ \ \ \ (VII.8)

\bigskip

To preserve clarity, we remember that in the earlier case, we recovered \ R.N.
metric from Schwarzschild's, by means of the transformation:

\bigskip

$-\frac{2m}{r}\rightarrow$\ \ $-\frac{2m}{r}+\frac{GQ^{2}}{c^{4}r^{2}}$
\ \ \ \ \ . \ \ \ \ \ \ \ \ \ \ \ \ \ \ \ \ \ \ \ \ \ \ \ \ \ \ \ \ \ \ \ \ \ \ \ \ \ \ \ \ \ \ \ \ \ \ \ \ \ \ \ \ \ \ \ \ \ \ \ \ \ \ \ \ \ \ \ \ \ \ \ \ \ \ \ \ \ \ \ \ \ \ \ \ \ \ \ (VII.9)

\bigskip

\bigskip

\bigskip{\Large VIII. Conclusion}

\bigskip

We have shown how to derive in a simple way, with a modest mathematical
apparatus, all known basic black hole metrics. The contents of this paper will
be a subject in the book by Berman (2007, to be published).

\bigskip

\bigskip\ \ {\Large Acknowledgements}

\bigskip I am grateful and thank my intellectual mentors, Fernando de Mello
Gomide and M.M. Som, and am also grateful for the encouragement by Geni,
Albert, and Paula. Marcelo Fermann Guimar\~{a}es, Antonio F. da F. Teixeira,
Nelson Suga, and others, contributed significantly towards the completion of
this paper.

\bigskip

{\Large References}

\bigskip

Adler, R.J.; Bazin, M.; Schiffer, M. (1975) - \textit{Introduction to General
Relativity}. 2nd Edition. McGraw-Hill. New York.

Berman, M.S. (2007) - \textit{A Primer in Black Holes, Mach's Principle and
Gravitational Energy. }To be published.

\bigskip Gruber, R.P.; Price, R.H.; Matthews, S.M.; Cordwell, W.R.; Wagner,
L.F. (1988) - American Journal of Physics \textbf{56}, 265.

Newman, E. T.; Couch, E.; Chinnapared, R.; Exton, A.; Prakash, A.; Torence, R.
(1965) - Journal of Mathematical Physics \textbf{6}, 918.

Kerr, R. P. (1963) - Physical Review Letters, \textbf{11}, 237.

\bigskip Rabinowitz, M. (2006) - Chapter 1 in \textit{Trends in Black Hole
Research} , Nova Science, New York.

Raine, D.; Thomas, E. (2005) - \textit{Black Holes - an Introduction }Imperial
College, London.

Reissner, H. (1916) - Ann. Phys. \textbf{50}, 106.

Schwarzschild, K. (1916) - Stizber. Deut. Akad. Wiss., Berlin, K1. Math.-Phys.
Tech., s. 189.

\bigskip Sexl, R.; Sexl, H. (1979) - \textit{White Dwarfs-Black Holes: An
Introduction to Relativistic Astrophysics} , Academic Press, New York.

Taylor, E.F.; Wheeler, J.A. (2000) - \textit{Exploring Black Holes -
Introduction to General Relativity}, Addison-Wesley Longman, San Francisco.

Thirring, H.; Lense, J. (1918) - Phys. Z. \textbf{19}, 156.

Winterberg, F. (2002) - \textit{The Planck Aether Hypothesis, }Gauss Scie.
Press, N.V.

\end{document}